# Deep Learning Based Segmentation of Various Brain Lesions for Radiosurgery

**Siang-Ruei Wu**
**Department of Electrical Engineering, National Taiwan University**

**Hao-Yun Chang**
**Department of Electrical Engineering, National Taiwan University**

**Florence T Su**
**College of Arts and Sciences , Santa Clara University**

**Heng-Chun Liao**
**School of Medicine, National Taiwan University**

**Wanju Tseng**
**Data Intelligence and Application Division, QNAP Systems, Inc**

**Chun-Chih Liao**
**Department of Neurosurgery , Taipei Hospital**

**Feipei Lai**
**Department of Computer Science & Information Engineering, National Taiwan University**

**Feng-Ming Hsu**
**Department of Oncology, National Taiwan University Hospital**

**Furen Xiao**
**Department of Neurosurgery, National Taiwan University Hospital**

## Abstract

Semantic segmentation of medical images with deep learning models is rapidly developed. In this study, we benchmarked state-of-the-art deep learning segmentation algorithms on our clinical stereotactic radiosurgery dataset, demonstrating the strengths and weaknesses of these algorithms in a fairly practical scenario. In particular, we compared the model performances with respect to their sampling method, model architecture, and the choice of loss functions, identifying the suitable settings for their applications and shedding light on the possible improvements.

Keywords: deep learning, image segmentation, brain tumors, radiosurgery, magnetic resonance imaging

# 1. Introduction

Stereotactic radiosurgery (SRS) is a treatment modality using ionizing radiation, focusing on precisely selected areas of tissue. It is usually delivered in a single session, but the radiation dose can also be fractionated. Targeting accuracy and anatomic precision are critical to successful SRS but are historically secondary concerns in other types of radiation therapy (Adler et al., 2004). Although SRS can be performed in many parts of the body, it is best-known to treat intracranial lesions. The common indications for intracranial SRS include many different types of brain tumors, vascular malformations (including arteriovenous malformation, AVM), and functional diseases such as trigeminal neuralgia (TN). Brain metastases, vestibular schwannomas, meningiomas, and pituitary adenomas are common tumor types treated by SRS.

Before the delivery of SRS to the target (e.g. a brain tumor), a detailed treatment planning begins with precise contouring of the target by a neurosurgeon or a radiation oncologist. The contouring is performed on computed tomography (CT) or magnetic resonance images (MRI). Sometimes, both CT and MRI are used, depending on devices and diseases. Normal organs or tissues sensitive to radiation are also contoured so that radiation dose and risk of injury can be estimated. These normal organs are called critical organs or organs at risk (OARs). In terms of image analysis, "precise" segmentation of targets and OARs is mandatory for SRS treatment planning. In current practice, the segmentation is manual and performed by professional personnel. The task is sometimes onerous and tedious, thus many research suggests computer assistance.

As convolutional neural networks (CNN), the dominant deep learning models, lead the breakthrough in computer vision recently, it also dominates MRI segmentation tasks. Havaei et al. (2017) proposed the idea of using a deep learning model to perform brain tumor segmentation tasks on MRI images. They pointed out that both local and global representations are essential to produce better results, and this intuition is later realized in various ways. Kamnitsas et al. (2017) later perfected this idea and achieved state-of-the-art performance with a two-path model. On the other hand, U-Net (Ronneberger et al., 2015, p.) was first proposed for the cell-tracking task, but then became widely used in many other segmentation tasks(Dong et al., 2017; Livne et al., 2019). In MICCAI BraTS 2017 competition (Bakas et al., 2017), most participants used U-Net variants, as the winner (Kamnitsas et al., 2018) simply ensembleed three kinds of the most common deep learning models, namely FCN(fully convolutional network, Shelhamer et al., 2017), V-Net(Milletari et al., 2016), and DeepMedic(Kamnitsas et al., 2017).

However, there were few studies applying deep learning methods to the actual SRS datasets (Liu et al., 2017). Different from the BraTS competitions, real applicable models may need to handle much more diversity rather than a single type of disease. We realized the necessity to explore the behavior in such a realistic scenario for the technology to achieve an applicable performance. Therefore, we collected a relatively large dataset with 1688 patients and analyzed the performance of models with various types of settings and architectures. More specifically, we benchmarked the performance of different segmentation models previously proposed on other tasks and also compare the effectiveness of various sampling methods and the choice of loss functions.

## 2. Material and methods

### 2.1 Dataset

#### 2.1.1 NTUH (National Taiwan University Hospital) dataset

The data was extracted from a medical center in northern Taiwan. The SRS device used was CyberKnife (Accuray, Sunnyvale, CA) and it has commenced operation since January 2008. In the decade until December 2017, there were 2578 treatment courses completed in 2411 patients. Among these, 2036 treatment courses of 1921 patients were intracranial.

We only selected patients undergoing first SRS with contrast-enhanced T1-weighted (T1+C) MRI image available. Finally, there were 1688 patients included in our dataset. Their data was arbitrarily divided into training and test set (Table 1). However, because treatment targets for patients with trigeminal neuralgia are not tumor nor vascular malformation, their data were all assigned to the training set.

|  | Train | Test |
| --- | --- | --- |
| Metastases | 504 | 53 |
| Meningioma | 314 | 29 |
| Schwannoma | 305 | 20 |
| Pituitary | 147 | 8 |
| AVM | 80 | 6 |
| TN | 38 | 0 |
| Other tumors | 169 | 15 |
| Total | 1557 | 131 |

Table 1. Clinical diagnoses of 1688 patients in the final dataset.

For each patient, the target was extracted from the treatment planning system, together with axial T1+C MRI (1 to 2 mm in slice thickness). There might be two or more targets in an image volume, which was particularly true in patients with brain metastases. After proper registration and de-identification, these image/label pairs were used for the training and evaluation of deep neural networks.

There were a total of 2568 distinct targets in these 1688 image sets. The target volumes ranged from 20 to 72646 $mm^3$, with a median of 1236 $mm^3$ and a mean of

3696±6637 mm$^3$. In 1013 image sets, there was only one target. The number of targets may range up to 34 in a single image set.

### 2.1.2 BraTS dataset

The BraTS 2015 dataset is a standard benchmark dataset for MRI segmentation tasks. It includes 220 multi-modal scans of patients with high-grade glioma (HGG) and 54 with low-grade glioma (LGG). T1-weighted, contrast-enhanced T1+C, T2-weighted and FLAIR images are available. The data has a common dimension of 240 x 240 x 155 with 1 mm$^3$ resolution. The annotation contains 5 classes: 0 for background, 1 for necrotic core (NC), 2 for oedema (OE), 3 for non-enhancing core, and 4 for enhancing core. The evaluation follows the rules of the competition by merging the predictions into three sets: whole tumor (classes 1,2,3,4), core (classes 1,3,4) and enhancing core (class 4).

## 2.2 Preprocessing

The raw data of the NTUH dataset contains images of different resolutions and fields of view (FOVs). In order to process them with the CNN, we first used the skull stripping function of Brain-Suite (Shattuck and Leahy, 2002) to locate the brain, then utilized the location of the brain mask to center and crop the MRI to the size of 200x200x200 mm$^3$. Lastly, we normalized them by the z-scores.

As for the BraTS dataset, it was already registered, cropped, and normalized with bias field corrections. We only normalized the data by the z-scores for every pulse sequence (T1, T2, T1+C, FLAIR).

## 2.3 Data Augmentation

To perform a fair comparison of model architectures, we established the following standard data augmentation in the training phase. For 2D models, we performed data augmentation with translation, rotation, shear, zoom, brightness, and elastic distortion (Dong et al., 2017). For 3D models, we did not perform any type of data augmentation.

## 2.4 Deep Learning Models

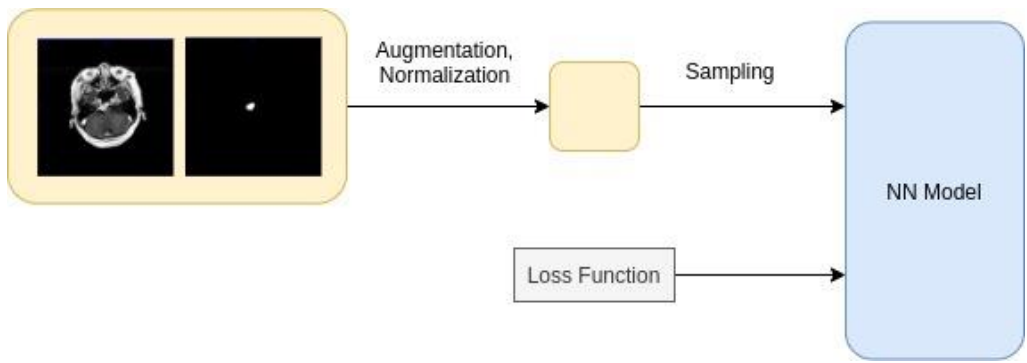

### 2.4.1 DeconvNet

DeconvNet (Noh et al., 2015) is an architecture adopted from VGG16, a 16-layered CNN by Visual Geometry Group, and is rather simple to implement. The objective of this design is to overcome the limitations of FCN, which cannot detect objects that are bigger or smaller than a specific size. In this case, the object may be fragmented or mislabeled. Besides, FCN only uses one convolution transpose layer to construct its output, so the output loses much detail. As a consequence, DeconvNet uses several layers of transpose-convolution and up-pooling.

The model can be divided into two parts: the encoder and the decoder, which are formed by convolution and deconvolution operations respectively. It is worth noting that we replaced the max-pooling and up-sampling operations by setting the stride of Conv and Deconv to 2 in our implementation. This is inspired by the recent proposition of generative adversarial networks.

### 2.4.2 DeepMedic

DeepMedic (Kamnitsas et al., 2017) is another kind of 3D CNN. It is special for taking two inputs, high resolution and low resolution. This design seeks to balance between fine structures and high-level information. Both inputs go through a series of convolution layers with skip structure, and then it constructs the output by fusing the features of both pathways.

### 2.4.3 PSPNet

Pyramid scene parsing network (Zhao et al., 2017), or PSPNet, is a state-of-the-art model in scene parsing tasks. We included it because it is also suitable for our segmentation task. The PSPNet utilizes the high-level representations extracted by a pretrained network and a novel design of the pyramid pooling module serves as a backend to predict the segmentations. The pyramid pooling modules pools the extracted feature maps in order to obtain features of different scales. The pretrained model is typically a ResNet trained on imagenet dataset [REF]. However, on an MRI dataset, the features are not transferable due to the large consistence and the absence of common pretrained models to process MRI images. In our implementation, we randomly initialize the ResNet backend and also remove the deep supervision loss.

### 2.4.4 U-Net

U-Net (Ronneberger et al., 2015) tries to improve the fine structure of segmentations and increase the amount of context used. Traditionally, when certain amount of pooling is required if one is intending to train with large patches, but unfortunately it degrades the performance as in FCN and DeconvNet. Hence, U-Net model utilizes skip connections to forward the unpooled features, thus the model can utilize the information of various scales. In our implementation, we abandoned the max-pooling and up-sampling operations for the same reason as in DeconvNet.

### 2.4.5 V-Net

V-Net (Milletari et al., 2016) is the adaption of U-Net for 3-dimensional data to capture the relationships in consecutive slices, which were omitted in the 2D models. It replaces the convolution and pooling operations with 3D versions.

## 2.5 Sampling Methods

### 2.5.1 two_dim

For two-dimensional models, we split the MRI data slice by slice and perform predictions separately. This may result in noises along the sliced axis due to the loss of spatial contiguity information.

### 2.5.2 three_dim

For three-dimensional models, the basic strategy is to feed the whole brain image data directly. While we experiment this setting on the BraTS2015 dataset, we found it causing overfitting and we suspect that it is because many of the voxels are irrelevant and redundant for the prediction. Thus we added two more three-dimensional sampling methods described below.

### 2.5.3 uniform_patch

To reduce the redundant voxels and save memory usage, we can sample small patches within the brain regions. While inferencing, we simply reassembled the patch predictions together.

### 2.5.4 center_patch

It was suggested that patches containing foreground regions are crucial to the training (Noh et al., 2015). We thus deployed this sampling strategy which guarantees at least one foreground voxel in the patch.

## 2.6 Loss Functions

Class imbalance is a major problem in most tumor segmentation problems, and it is even more severe in our task compared to the BraTS glioma dataset because of small target volumes. The imbalance would most likely lead the model to trivial solution, which predicts all voxels as background. There are several ways to deal with this problem by modifying the loss function.

### 2.6.1 weighted-cross-entropy

Re-weighting the sparse class is the most common solution to the class imbalance problem. In this study, we set the class weights inversely proportional to the ratio of the class. In particular:

$C = -\sum_{c=1}^{M} g_{oc} log(p_{oc}) \times \frac{1}{r_c}$ where M is the number of classes and $r_c$ is the ratio of class c in the whole volume / dataset (as an implementation choice), $g_{oc}$ is the ground-truth label of a voxel and $p_{oc}$ is the predicted label probability of a voxel of class c.

### 2.6.2 soft-dice

In (Milletari et al., 2016), authors suggest using the differentiable version of dice-score, namely soft-dice, directly as the objective due to its resistibility to class imbalance. It is fairly natural to use this loss function because the dice-score is the most common evaluation metric in related tasks.

There are two implementations of the soft-dice loss function. Regarding the cardinality of sets, one can perform summation directly or with squaring. In particular:

$$D_1 = \frac{2\sum_i^N p_i g_i}{\sum_i^N p_i^2 + \sum_i^N g_i^2} \quad \text{or} \quad D_2 = \frac{2\sum_i^N p_i g_i}{\sum_i^N p_i + \sum_i^N g_i}$$

where $p_i$ is the predicted label probability and $g_i$ is the ground-truth label. We found the two versions producing almost identical performances. In this study, we refer to the second version as the soft-dice loss function.

# 3. Results

Some representative results were shown in Fig.1. The overall dice scores of these networks on the NTUH dataset ranged from 0.33 (DeepMedic) to 0.51 (V-Net). Table 2 showed the detailed performance of each network.

1. Prediction with low dice score

| ground truth | deconvnet | deepmedic |
|---|---|---|
| 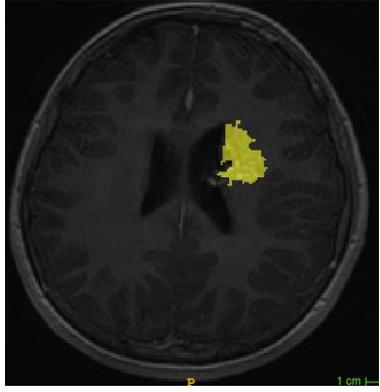 | 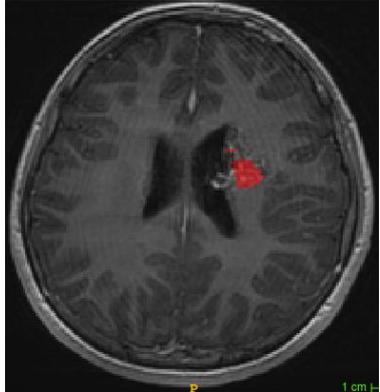 | 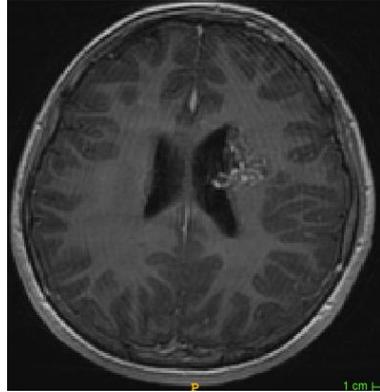 |
| pspnet | u_net | v_net |
| 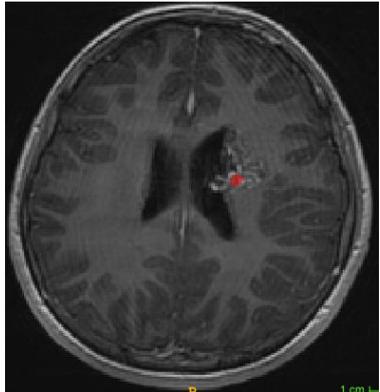 | 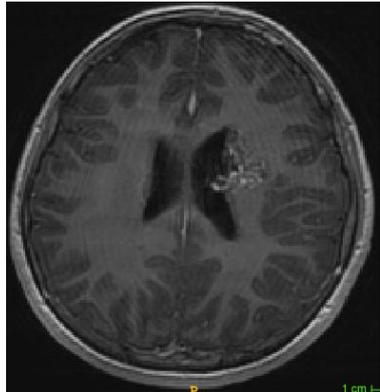 | 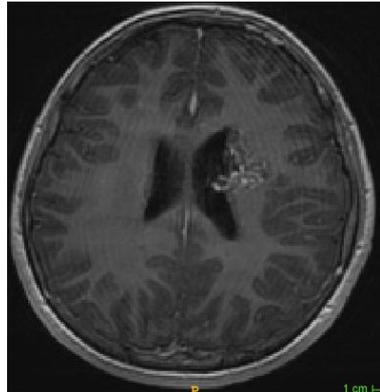 |

2. Predict with average dice score

| ground truth | deconvnet | deepmedic |
|---|---|---|
| 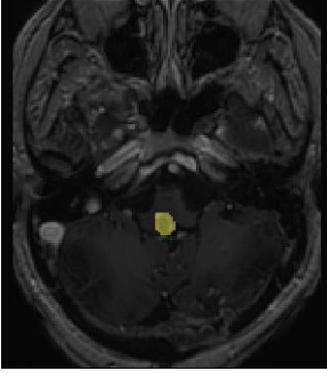 | 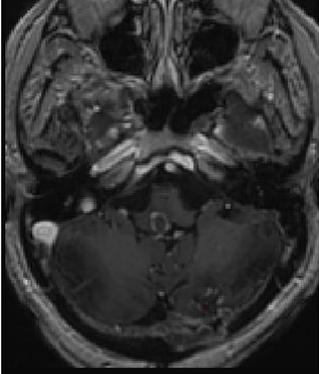 | 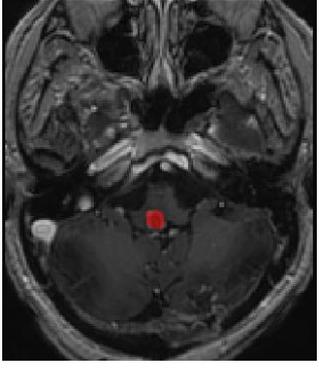 |
| pspnet | u_net | v_net |
| 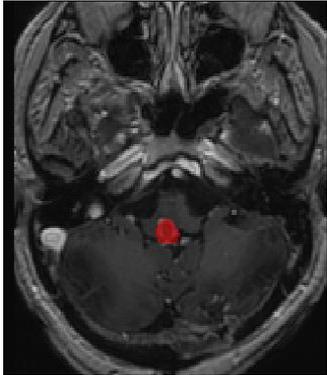 | 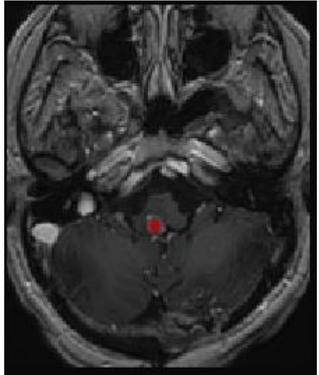 | 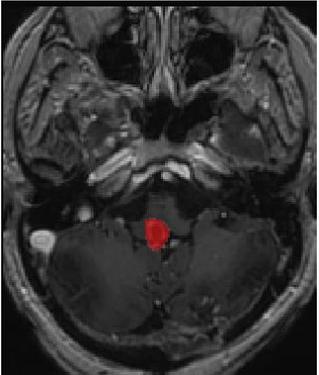 |

3. Prediction with high dice score

| ground truth | deconvnet | deepmedic |
|---|---|---|
| 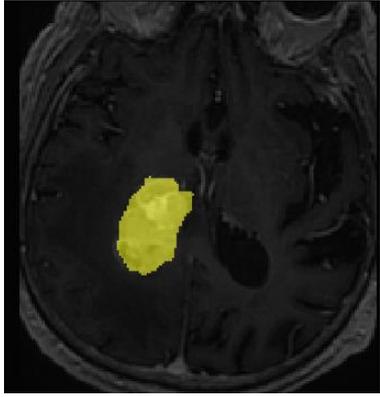 | 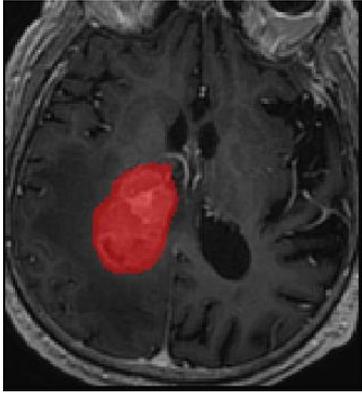 | 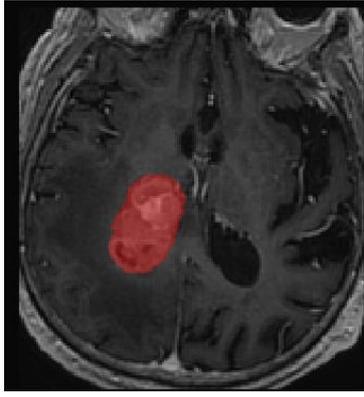 |
| pspnet | u_net | v_net |

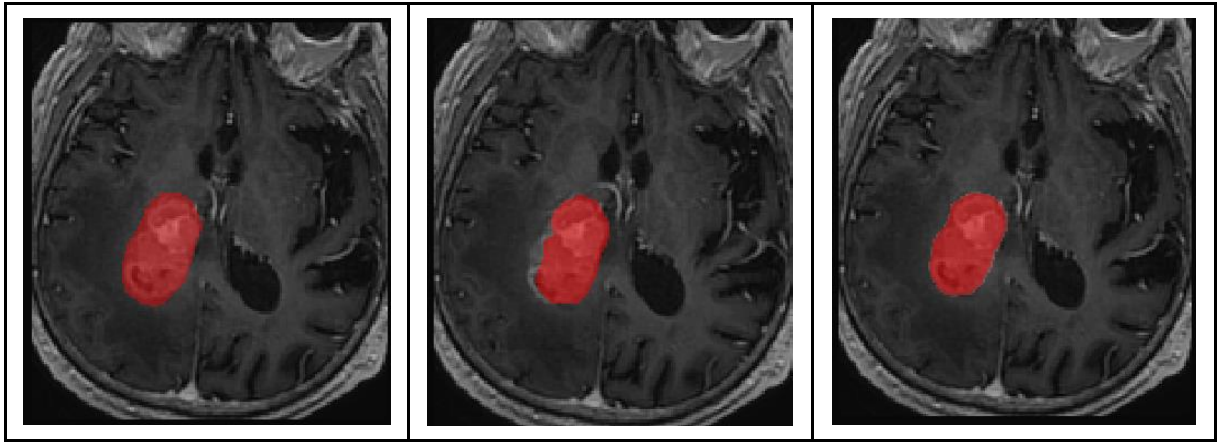

Fig. 1. Three cases from the NTUH dataset, showing representative results when using different models.

## 3.1 Performance comparison

NTUH

| model | num parameters | batch sampler | loss function | val precision | val sensitivity | val hard-dice |
|---|---|---|---|---|---|---|
| deconvnet_big | 12544324 | two_dim | ce_minus_log_dice | 0.46 | 0.48 | 0.43 |
| u_net | 34524034 | two_dim | ce_minus_log_dice | **0.48** | 0.48 | 0.43 |
| pspnet_2d | 28280773 | two_dim | ce_minus_log_dice | 0.47 | 0.48 | 0.43 |
| v_net | 8232274 | uniform_patch3d | ce_minus_log_dice | 0.39 | 0.54 | 0.41 |
| v_net | 8232274 | three_dim | cross entropy | 0.2 | 0.56 | 0.25 |
| v_net | 8232274 | three_dim | ce_minus_log_dice | **0.48** | 0.51 | 0.46 |
| v_net_dropout0.1 | 8232274 | three_dim | ce_minus_log_dice | 0.47 | **0.66** | **0.51** |
| deepmedic | 1301478 | center_patch3d | ce_minus_log_dice | 0.36 | 0.43 | 0.35 |
| deepmedic | 1301478 | center_patch3d | cross entropy | 0.37 | 0.43 | 0.33 |

Table 2. Performance of different models on NTUH dataset.

BraTS

| | | | val-hard-dice | | |
|---|---|---|---|---|---|
| model | batch-sampler | loss_function | whole | core | enhancing |
| v_net | center_patch | crossentropy | 0.85 | 0.79 | 0.82 |
| v_net | center_patch | ce_minus_log_dice | 0.86 | 0.77 | 0.7 |
| v_net | three_dim | ce_minus_log_dice | 0.74 | 0.74 | 0.77 |
| v_net | uniform_patch | crossentropy | 0.77 | 0.72 | 0.73 |
| u_net | two_dim | ce_minus_log_dice | **0.87** | **0.83** | **0.83** |
| pspnet_resnet34 | two_dim | ce_minus_log_dice | **0.87** | 0.76 | 0.69 |
| pspnet_resnet50 | two_dim | ce_minus_log_dice | **0.87** | 0.79 | 0.74 |
| deepmedic | center_patch | crossentropy | 0.83 | 0.76 | 0.81 |

Table 3. Performance of different models on BraTS dataset.

For comparison, the performance of each network on the BraTS dataset was shown in Table 3. Compared to NTUH datasets, every network performed much better without exception. ce_minus_log_dice denotes cross-entropy - log(soft-dice).

On NTUH datasets, the performance was also affected by the types of lesions. We got better results for brain metastases, meningiomas, and schwannomas, while all models performed poorly on pituitary tumors and AVMs (Fig. 2).

Because of the nature of PSPnet and DeepMedic, they took a significantly longer time for inference (Table 4.). V-Net had the least number of parameters and the shortest inference time. We also find that adding dropouts in V-Net further improves its performance, which we have noted in the table with 0.1 being the dropout rate.

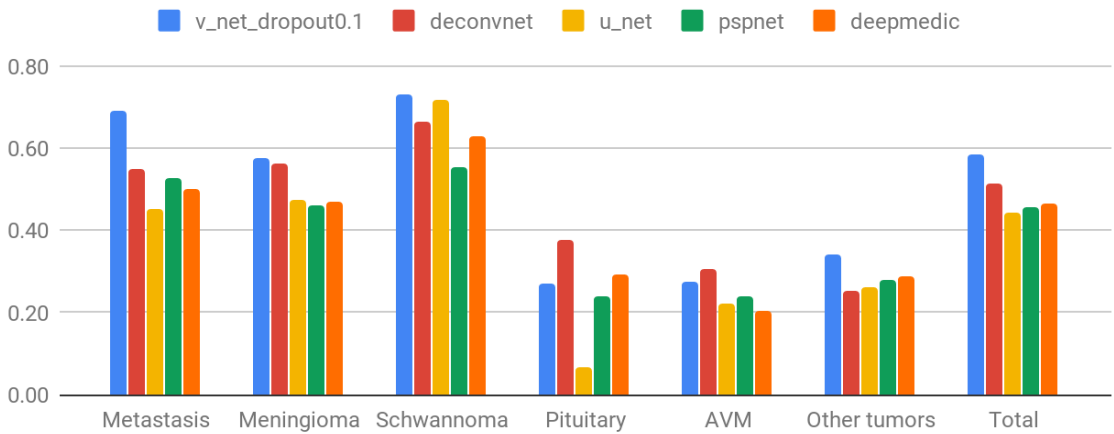

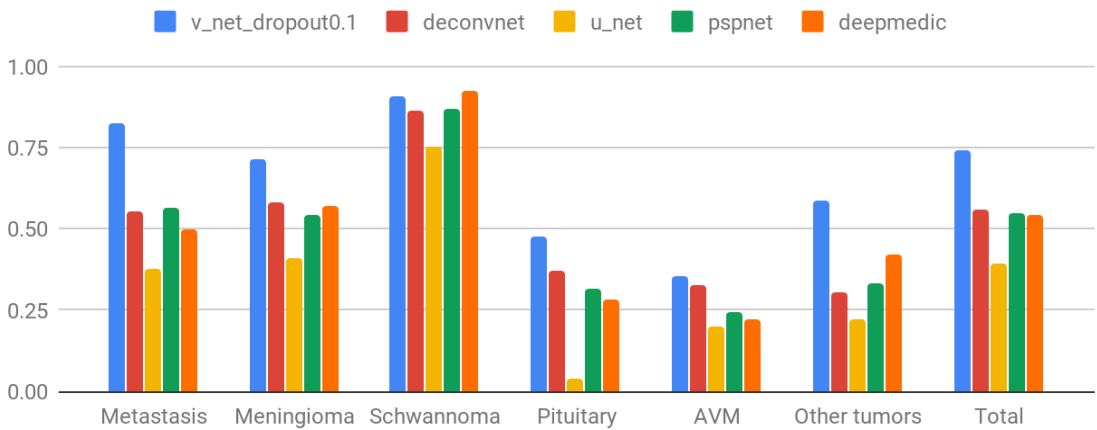

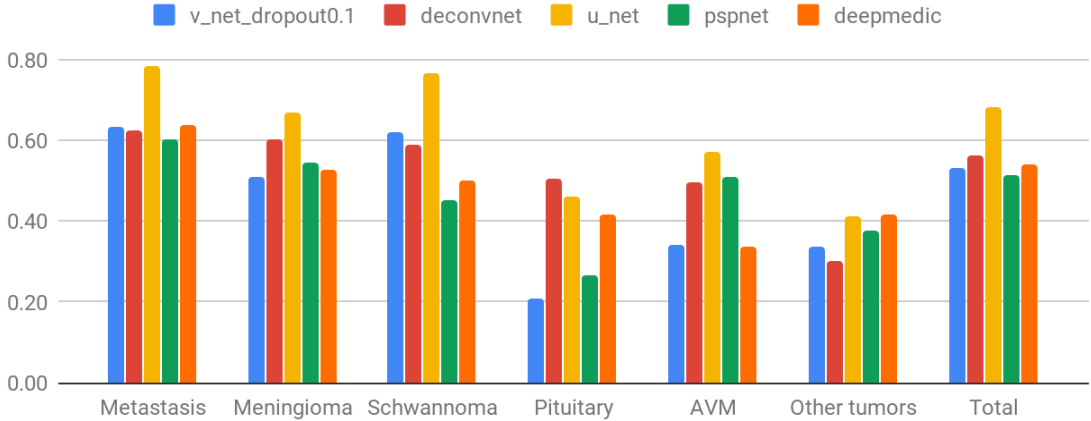

Fig. 2. Performance on different lesion types of NTUH dataset.

|  | v_net_dropout0.1 | deconvnet | u_net | pspnet | deepmedic |
|---|---|---|---|---|---|
| Inference time (minutes:seconds) | 02:51 | 04:00 | 04:01 | 14:43 | 17:13 |
| Number of parameters | 8.23M | 12.5M | 34.5M | 28.3M | 13M |

Table 4. Inference time on our hardware and parameters of different networks

## 4. Discussion
### 4.1 Segmentation performance: NTUH vs BraTS dataset

The performance on our radiosurgery dataset is far worse than that on BraTS. Many factors might lead to such a result. First of all, the tumor volumes in the NTUH dataset are typically smaller than those in BraTS 2015. On average, the tumor occupied 1.23% of the whole image volume in the BraTS dataset, but only 0.145% in ours. It should also be noted that a significant portion of our dataset contained multiple targets, which is much less likely for glioma patients (BraTS). The lesions in NTUH dataset are thus more difficult to detect.

Moreover, there is significant heterogeneity in our dataset. In contrast to BraTS containing only glioma cases, our dataset includes cranial tumors of various pathology. In a strict sense, we also had some images of non-neoplastic diseases like AVM. Because many of these tumors are actually extra-axial (outside the brain parenchyma) and may even extend extracranially, we can't perform skull stripping like BraTS. Because of the heterogeneity of tumor types and sites, we may need a much larger dataset to reach similar performance.

Another reason is that we only used one image set (T1+C) to predict instead of four sequences used in the BraTS dataset. Less information might lead to deteriorated performance.

It is also worth mentioning that our dataset is quite imbalanced disease-wise. From the performance of the models we've trained, we can observe that this imbalance results in serious bias issues for the minority patients. We found it quite difficult to train a model by the traditional soft-dice loss or cross-entropy loss. Using the weighted cross-entropy loss gives us a 0.25 dice score while our modification of subtracting a log-soft-dice term improves the dice score to 0.40. Such difference may result from tumor size, since tumors in our dataset had of fewer voxels on average. In addition to the data variety, the weighted-cross-entropy function could be very unstable and thus harmful to the optimization. Empirically we found that it's most likely the model will fail in 10 epochs and predict nothing but the background for all inputs. By adding another term with the dice score, the new loss function provides better guidance to the model, and we can empirically observe the significant improvements.

Although the targets in our dataset were defined and contoured by experienced clinicians, it should be noted that they were the targets we wanted to treat. Therefore, not every lesion detected by human experts was labeled. For example, it is very possible that a patient with brain metastases also has another meningioma, which may be stable and will not be labeled and treated by radiosurgery. If an algorithm detects that meningioma, decreased precision and dice score can be expected.

## 4.2 Performance on different types of tumor

We can see that these models performed better for brain metastases, meningiomas, and schwannomas, where there are more than 300 cases each. They performed best for schwannomas, probably because most of these are vestibular ones, whose locations are always around internal auditory meatus.

On the other hand, these models performed poorly for pituitary tumors, AVMs and other tumor types. Besides the relatively small number of cases for training, pituitary tumors and AVMs are not always readily visible for humans using only T1+C series. For example, dynamic contrast-enhanced MRI may be required to visualize pituitary tumors. AVMs are sometimes not visible even using time-of-flight (TOF) MRI, so computed tomography angiography / digital subtraction angiography may be required for target contouring.

## 4.3 Comparison between deep learning models

With respect to the input format, there are two classes of model architectures. 2D model predicts tumor in just one slice and completely discards the information along the z-axis, while 3D model utilizes the full information on the MRI volume. This results in a trade-off between features and overfitting. When receiving more features, it is more likely to overfit the unrelated noise, especially with such a small dataset. As a result, proposed methods often restrain the receptive field and predicts on patches of inputs typically 64x64x64 mm$^3$. We examined this trade-off in our benchmark experiment on the BraTS dataset. Surprisingly, when experimenting V-Net on our dataset, small patch-wise prediction becomes detrimental and receiving the full brain volume results in the best performance.

Overall, 3D models seem to be more appealing. 3D models present the full potential of convolution networks, reducing the number of parameters and becoming far more efficient due to its convolution nature. Specifically, V-Net has approximately 1/30 of parameters compared to U-Net, shortest inference time, and the best performance on dice metric. The only shortcoming of 3D models is the requirement of GPU RAM due to the large input. In our experiments, we solve this by using smaller batch-size. Furthermore, replacing batch normalization with dropout is quite effective in preventing overfitting because of the small batch size.

## 5. Conclusion

We benchmarked 5 commonly used deep learning segmentation models on our SRS dataset. We confirmed that these approaches also work on a heterogeneous dataset, with decreased performance. We discovered that the V-Net architecture worked best for this specific task. With top dice scores, smallest size of the model, and shortest inference time, V-Net may be a good choice to improve upon. We also found that when training on the dataset with such heterogeneity and class imbalance, using weighted cross-entropy loss with log-soft-dice term significantly improves the performance.

## Acknowledgments

This work was supported by the Ministry of Science and Technology, Taiwan, ROC [grant numbers 107-2634-F-002-015, 108-2634-F-002-015].

# Supplementary Material

1. Github Repository: https://github.com/raywu0123/Brain-Tumor-Segmentation
2. Detail of NTUH experiment

| DICE | v_net_dropout0.1 | deconvnet | u_net | pspnet | deepmedic |
|---|---|---|---|---|---|
| Metastasis | **0.69** | 0.55 | 0.45 | 0.52 | 0.50 |
| Meningioma | **0.57** | 0.56 | 0.48 | 0.46 | 0.47 |
| Schwannoma | **0.73** | 0.66 | 0.72 | 0.55 | 0.63 |
| Pituitary | 0.27 | **0.38** | 0.07 | 0.24 | 0.29 |
| AVM | 0.27 | **0.31** | 0.22 | 0.24 | 0.20 |
| Other tumors | **0.34** | 0.25 | 0.26 | 0.28 | 0.29 |
| Total | **0.59** | 0.52 | 0.44 | 0.46 | 0.46 |

| SENSITIVITY | v_net_dropout0.1 | deconvnet | u_net | pspnet | deepmedic |
|---|---|---|---|---|---|
| Metastasis | **0.82** | 0.55 | 0.37 | 0.56 | 0.50 |
| Meningioma | **0.71** | 0.58 | 0.41 | 0.54 | 0.57 |
| Schwannoma | 0.91 | 0.86 | 0.75 | 0.87 | **0.92** |
| Pituitary | **0.48** | 0.37 | 0.04 | 0.31 | 0.29 |
| AVM | **0.36** | 0.33 | 0.20 | 0.24 | 0.22 |
| Other tumors | **0.58** | 0.30 | 0.22 | 0.33 | 0.42 |
| Total | **0.74** | 0.56 | 0.39 | 0.55 | 0.54 |

| PRECISION | v_net_dropout0.1 | deconvnet | u_net | pspnet | deepmedic |
|---|---|---|---|---|---|
| Metastasis | 0.63 | 0.62 | 0.78 | 0.60 | **0.64** |
| Meningioma | 0.51 | 0.60 | **0.67** | 0.54 | 0.53 |
| Schwannoma | 0.62 | 0.59 | **0.76** | 0.45 | 0.50 |
| Pituitary | 0.21 | **0.50** | 0.46 | 0.27 | 0.42 |
| AVM | 0.34 | 0.49 | **0.57** | 0.51 | 0.34 |
| Other tumors | 0.34 | 0.30 | **0.41** | 0.38 | 0.42 |
| Total | 0.53 | 0.56 | **0.68** | 0.52 | 0.54 |